**Modeling pest dynamics in trap cropping to improve yield: the effects of attraction, retention, and land allocation**

Matthew H Holden

The University of Queensland

## Abstract

Trap crops reduce damage to a cash (main) crop by attracting pests away from it. Yet this protection is weakened when pests disperse back into the cash crop. In this paper, we focus on the importance of preventing this backflow, showing that effective trap cropping depends jointly on how strongly pests are attracted to trap plants and how rarely they leave them. Together with the proportion of the field devoted to trap plants, these processes determine both the efficacy and feasibility of trap cropping at commercial scales. We formalise this relationship using a simple yield-maximisation framework, in which growers weigh pest suppression benefits against the land sacrificed to trap plants. The model shows that when dispersal from trap plants equals that from the cash crop, optimal trap coverage can exceed 20 –30 % of the landscape, levels rarely acceptable to growers. However, reducing pest dispersal off trap plants to just one-quarter of cash crop dispersal lowers the optimal required trap area to ~5 %, transforming trap cropping from impractical to feasible. Understanding these relationships can guide trap-cropping design, from plant choice to targeted interventions that reduce pest movement, to minimise damage, maximise yield, and make trap cropping a reliable component of sustainable pest management.

## Introduction

Trap cropping, the use of alternative host plants to draw insect pests away from a focal cash crop, has long been promoted as a promising strategy for sustainable pest management. In theory, it offers growers a way to reduce pesticide use while maintaining yield, and it has been studied in hundreds of agricultural systems (Hokkanen 1991; Shelton & Badenes-Perez 2006; Badenes-Pérez & Hokkanen 2024). Yet despite this conceptual appeal, reviews of more than a hundred systems found only a few handfuls of consistent successes, most requiring additional inputs such as insecticide applications to trap plants (Shelton & Badenes-Perez 2006; Holden 2012, Sarkar et al 2018). The puzzle is clear: why does a strategy with such strong appeal so often fail in practice, and how can we increase its chance of success?

Mathematical and simulation models provide one way forward because they allow researchers to test scenarios at scales infeasible in the field. Early modelling efforts established that trap cropping could, in theory, suppress pest populations, under specific conditions. Models of herbivore movement across fields (Banks & Ekbom 1999; Hannunen 2005) showed that suppression improved with both the strength of attraction and the proportion of the field devoted to trap crops. Behavioural extensions incorporated host-finding and settling behaviour and examined alternative spatial layouts (e.g. borders, rows, or patches) but still relied on high trap-crop coverage to achieve success (Potting, Perry & Powell 2005). In these models, trap cropping appeared promising, possibly because trap plants occupied a substantial fraction of the simulated landscape (so that pests leaving one trap plant were likely to land on another rather than return to the cash crop). Fenoglio et al. (2017) made this dependence explicit: when trap crops covered less than 10 % of the field and pests reproduced freely, suppression was negligible unless additional control measures, such as pesticides on the trap plants, were introduced.

Subsequent work explored other more ecological forms of control. Banks & Laubmeier (2024) coupled pest and natural-enemy movement, demonstrating that suppression improved dramatically when predators or parasitoids tracked pest aggregations on trap plants.

Viewed together, these studies chart a clear progression. As models incorporated greater biological realism, from spatial arrangement to behaviour and enemy interactions, the same limitation reappeared: trap cropping only worked when pest movement from trap plants back to the cash crop was curtailed. Whether through insecticides, natural enemies, or extensive trap-crop area, the systems that succeeded were those that effectively prevented pests from returning to the main crop.

Holden et al. (2012) formalised this mechanism by explicitly modelling pest movement between trap and cash crops. Their framework showed that reducing the rate at which pests leave trap plants can sharply lower pest densities on the cash crop. Further, their review of empirical systems supported this pattern: all successful large-scale cases combined attraction with some intervention that curtailed pest movement off trap plants, whether through pesticide treatments (Srinivasan & Krishna Moorthy 1991; Dogramaci et al. 2004; Cavanagh et al. 2009; Lu et al. 2009), physical removal of pests (Swezey, Nieto & Bryer 2014), sticky traps (Moreau & Isman 2011), or planting trap crops at very high densities (Michaud, Qureshi & Grant 2007).

Preventing dispersal of pests back into the cash crop is now increasingly recognised as a critical design element of successful trap cropping (Blaauw et al. 2017). Whereas early empirical work largely emphasised attraction (Palaniswamy, Gillott, & Slater 1986; Lu et al 2004; Badenes-Pérez et al 2004; Lee, Nyrop & Sanderson 2009), more recent studies have begun to measure pest *arrestment*, the tendency of pests to remain on trap plants rather than disperse away, alongside attraction (Boiteau et al. 2014; Walsh et al. 2016; Blaauw et al. 2017; Wallingford et al. 2018; Owusu 2021). Despite this progress, the underlying principle that effective trap cropping depends on limiting dispersal from trap plants remains underexplored as the central focus of a single work.

In this chapter, we address this gap. We re-examine and simplify the Holden et al. (2012) model, but express results directly in terms of dispersal probabilities rather than their complements (sometimes called "retention"), since dispersal more naturally represents pest movement between plant types. This formulation shows how the relative rates of pest departure from trap and cash plants, together with the cash-to-trap planting ratio, jointly determine pest suppression. We then extend the framework to include yield optimisation, allowing growers to weigh pest-control benefits against the land costs of trap crops. The analysis reveals that reducing dispersal from trap plants can lower the required proportion of trap crops from impractical levels (≈25 %) to realistic ones (≈5 %), clarifying when trap cropping can be both effective and commercially feasible.

### A simple model of pest dynamics between cash and trap crops

We study the following model (see Table 1 for a list of parameters and variables), where $x_t$ is the proportion of the pest population on the main crop at time *t*. Therefore, $1 - x_t$, is the proportion of the pest population on the trap crop at time $t$. Note that throughout this chapter, we use the terms "cash crop" and "main crop" interchangeably to refer to the focal crop that trap plants are intended to protect. Pests leave main crop plants at each time step with probability $l$, and leave trap plants with probability $\lambda$. Pests leaving from a main crop plant settle on a main crop plant with probability, $s$, and pests leaving a trap plant settle on a main crop plant with probability, $\sigma$. Under these assumptions, the proportion of the pest population on the main crop changes through time iteratively according to the difference equation,

$$x_{t+1} = (1-l)x_t + lsx_t + \lambda\sigma\,(1-x_t). \tag{1}$$

The first term, $(1-l)x_t$, is the pests that stay on the previous main crop host plant. The second term, $lsx_t$, captures pests that leave and then resettle on the main crop, either settling on the same main crop plant or a different one. The third term, $\lambda\sigma\,(1-x_t)$, is influx from trap plants, pests that leave trap plants and then settle on the main crop.

**The effect of preventing dispersal off trap plants on long run pest densities**

The equilibrium of the model described in equation (1) is

$$x^* = \frac{\lambda\sigma}{\lambda\sigma + l(1-s)}. \tag{2}$$

This is approximately the proportion of the pest population on the main crop in the long run. Another useful quantity to consider is the long-term ratio of pests on the trap crop relative to the cash crop, $R$, which is,

$$R = \frac{1-x^*}{x^*} = \frac{l(1-s)}{\lambda\sigma}. \tag{3}$$

From equation (3), the long-run odds depend on two composite flows: backflow from trap plants to the main crop $\lambda\sigma$, and the flow off the main crop to the trap crop, $l(1-s)$.

For example, consider a field where ten percent of pests on the trap crop flow back to the cash crop in a single time step, $\lambda\sigma = 0.1$, and 50 percent of pests on the cash crop are drawn to and settle on the trap crop in the next time step, $l(1 - s) = 0.5$. Then, equation (3) says that it is five times more likely that any single pest individual will be on the trap plant rather than a main crop plant. This example shows that the balance of flows back and forth between cash plants and trap plants is the key driver of pest density on the cash crop.

In the following analyses, we rely on the above expressions for equilibrium pest densities and odds ratios as proxies for cumulative pest load, following Holden et al. (2012, 2026). The numerical simulations in Holden et al. (2012), and analytic derivations in Holden (2026), showed that equilibrium is reached rapidly under typical pest movement patterns, making this a reasonable simplification for systems where pests move repeatedly within a growing season.

So far, our simple model has assumed very little about the spatial structure of the field and how pests move across it. The only major assumption is that all trap plants are treated identical to one another, and the same for cash plants. This is critical because we assume all pests leaving trap plants have the same probability of settling on a cash plant, regardless of which trap plant they came from. The assumption likely holds in several systems. If trap plants are planted in rows, for example, from the eyes of a pest on a trap plant, locally, the world looks the same. At this stage, we have not yet specified a particular movement model. One would need to do this to determine the settlement probabilities on the cash plant when pests leave the trap plant, $\sigma$, and when they leave the cash plant $s$.

To propose an expression for these parameters, let us first assume the simplest possible movement model, often referred to as common pool dispersal, where pests have access to all plants in the field in a single time step. Let $n_T$ be the number of trap plants in the field, and $n_C$ be the number of cash plants. Also assume a trap plant is $a$ times more attractive than a single cash plant. At each time step, a pest that has left its current host plant chooses among all plants, making its previous location irrelevant. Therefore, the probability it settles on a cash plant is

$$s = \frac{n_C}{a n_T + n_C} = \sigma. \tag{4}$$

Assuming there are $n$ plants total in the landscape, then the number of cash plants is simply, $n - n_T$. Substituting this and expression (4) into (2) gives,

$$x^* = \frac{\lambda n_C}{a l n_T + \lambda n_C},$$

and the odds ratio an individual pest is on the trap relative to the cash crop is,

$$R = a\,\frac{l}{\lambda}\,\frac{n_T}{n_C}.$$

It also follows that, the equilibrium depends on parameters only through $R$,

$$x^* = \frac{1}{R+1}.$$

This means that the odds ratio, $R$, and equilibrium proportion of pests on the cash crop, $x^*$, are uniquely determined by the product of attractiveness, the ratio of the dispersal probabilities off main crop and trap crop, and the ratio of the number of trap plants to cash plants in the field. This highlights that attraction and relative dispersal act as equally powerful levers.

Suppose the grower would like to achieve a target odds ratio $R$ (ratio of pests on the trap crop to pests on the cash crop). The critical number of trap plants needed follows,

$$n_T = \frac{\lambda R}{a l} n_C.$$

This implies that effective control often requires a grower to dedicate a substantial share of their field to trap plants. For example, consider a trap crop which is five times more attractive than the cash crop, $a = 5$, but where pests are equally likely to leave a plant regardless of which type it is, $\lambda = l$, and suppose the grower would like twice as many pests on the trap crop as the cash crop, to avoid pest damage. In such a case the grower would need two trap plants for every five cash plants, which is nearly 30% of their field allocated to trap plants. It is clear from this exercise just how important it is to arrest pests on the trap crop, preventing dispersal back into the main field.

A particularly intuitive way to frame this for growers is in terms of the ratio of cash plants to trap plants, $n_C/n_T$. While probabilities such as λ, σ, and s describe the underlying pest dynamics, a grower is more likely to think in terms of 'how many rows of my cash crop do I lose if I add trap crops?' This makes the trade-off immediately tangible: a higher ratio means more land devoted to profitable production, while a lower ratio means a larger sacrifice of cash-crop area. Because adoption depends not only on biological efficacy but also on whether the required trap proportion is acceptable, this ratio connects the model outputs directly to grower decision-making.

In this spirit, the above conditions can be expressed as a critical cash to trap plant ratio, $n_C/n_T$,

$$\frac{n_C}{n_T} = \frac{al}{R\lambda}.$$

The grower would prefer the ratio to be high. That is, they want as many cash plants producing product as possible given a desired level of pest suppression. What is immediately clear is that the trap plant must be either incredibly attractive or reduce dispersal substantially relative to the cash plant for the ratio to be high. For example, if the grower wants to maintain at least 10 cash plants for every trap plant in their field and wants the odds of an individual pest being on the trap crop to be 10 times greater than on the cash crop, then they need $al/\lambda = 100$. This can be achieved by the leaving probability on the trap crop, $\lambda$, being 100 times lower than for the cash plant, $l$, or by a trap crop being 100 times more attractive. This trade-off between attractiveness and reduced dispersal is illustrated in Figure 1.

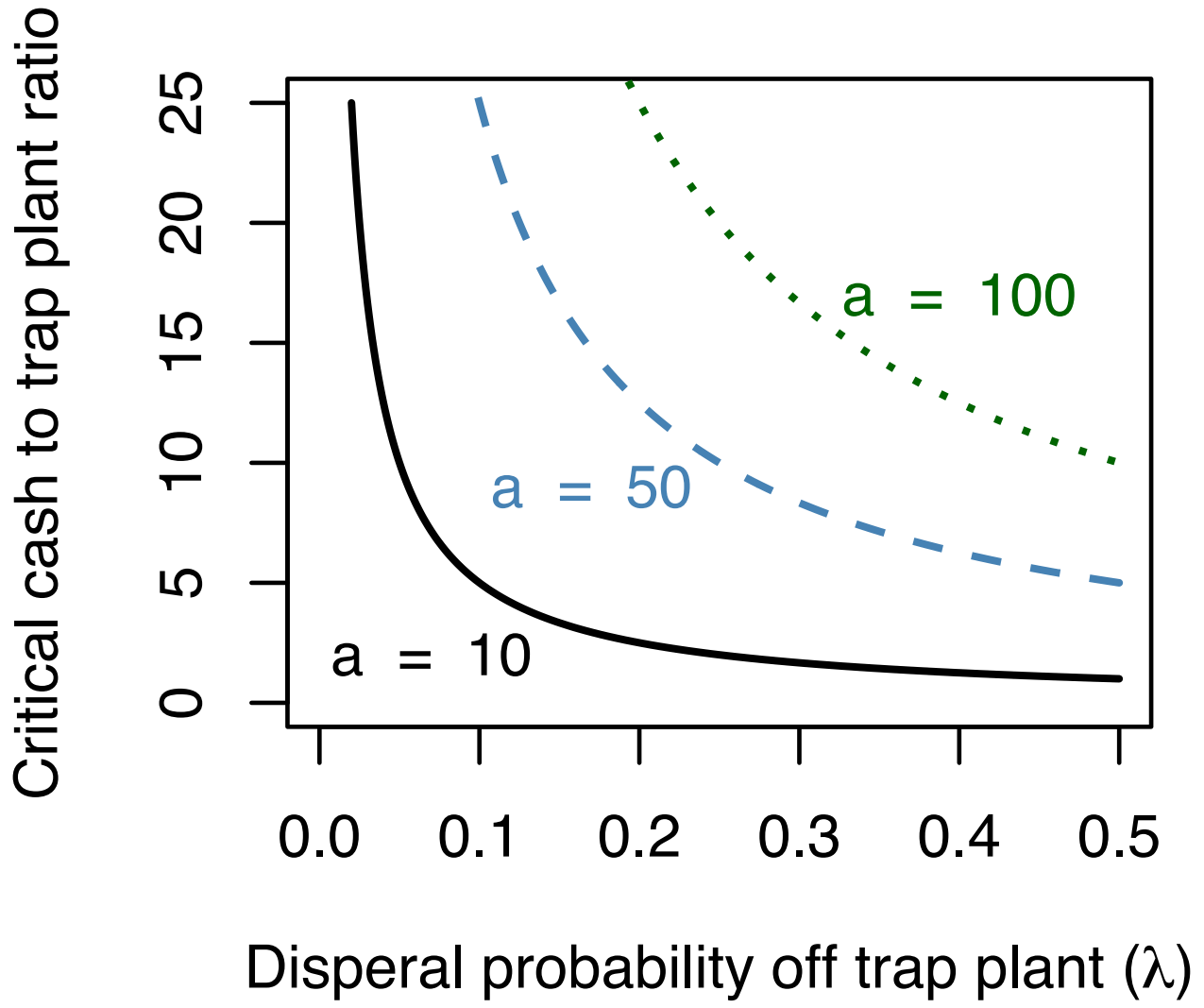

**Figure 1.** Critical cash-to-trap plant ratio ($n_C/n_T$) as a function of the dispersal probability off trap plants (λ), for different levels of trap crop attractiveness (a). Curves are shown for a = 10 (solid black), a = 50 (dashed blue), and a = 100 (dotted green). Parameter values are $l = 0.5$ and R = 10. A higher ratio indicates more cash plants can be supported per trap plant while still achieving the target pest suppression ratio. The figure shows that the curves for a = 10 and a = 50 only exceed a ratio of 10 when λ is very low, meaning dispersal from trap plants must be strongly reduced for trap cropping to be feasible. Only in the case of very high attractiveness (a = 100) can higher cash-to-trap ratios be maintained without such strong dispersal control.

This makes it clear that without reduced dispersal off of the trap crop, highly attractive trap plants can fail to protect the main crop. In practice, this means that growers cannot rely on attractiveness alone; preventing dispersal is a decisive factor that determines whether trap cropping is a viable pest management strategy.

**Table 1.** List of parameters and variables in the model

| Variable or parameter | Description |
|---|---|
| $x_t$ | Proportion of pests on the cash crop |
| $s$ | Settlement probability onto a cash plant given it left from a *cash* plant |
| $\sigma$ | Settlement probability onto a cash plant given it left from a *trap* plant |
| $l$ | Pest leaving probability off of a cash plant |
| $\lambda$ | Pest leaving probability off of a trap plant |
| $a$ | Attractiveness ratio: how many times more attractive a trap plant is relative to a cash plant. |
| $n_T$ | Number of trap plants |
| $n_C$ | Number of cash plants |
| $R$ | Odds ratio of a given pest being on a trap plant vs cash plant (e.g. $R$ =2 means pests are twice as likely to be on the trap plant) |
| $N$ | Total number of pests in the field or greenhouse |
| $x^*$ | Equilibrium (long run) proportion of pests on the cash crop |
| $r$ | Proportional reduction in the leaving probability from trap plants, meaning $\lambda = rl$, or a pest is $r$ times more likely to leave a cash plant than a trap plant |
| $\beta$ | Proportion of yield lost from pests in the absence of trap cropping (a.k.a the yield at risk) |

It should be noted that because the model our quite general, it can apply across different spatial scales. We formulated the model for a field or greenhouse, but it could equally apply at larger scales such as a farm or landscape provided the above assumptions are met for these scales. Throughout the paper we use field, farm, and landscape interchangeably to mean the managed spatial unit of interest.

**What is the optimal number of trap plants to maximize grower profits?**

The above analysis showed that effective pest suppression often requires a grower to devote a substantial share of their land to trap crops. However, growers are often reluctant to sacrifice this much space unless the yield benefits are obvious. Every trap plant displaces a profitable cash plant, creating a clear spatial trade-off: more trap plants mean fewer cash plants. The key question is whether the reduction in pest damage offsets the lost production area. In this section, we analyze this trade-off to identify the optimal proportion of the landscape to devote to trap plants, one that maximizes yield and therefore, grower profit. We also show how reducing dispersal from trap plants can shift this optimum, lowering the amount of land required for trap crops and making trap cropping more feasible for growers.

*The basic optimization model*

Consider a landscape with $n$ plants. Of these, $n_T$ are trap plants, so we have $n_C = n - n_T$ cash plants. In the simplest case, assume pest damage reduces yield in direct proportion to pest density. If pests are left unchecked, the cash crop produces zero yield; if no pests are present, the crop achieves its maximum, pest-free yield. As an equation, the above description means the yield of a plant is simply,

$$k\left(1 - \frac{x^* n}{n_C}\right),$$

where $k$ is the maximum, pest-free yield produced by an individual cash plant.

This framing immediately shows why some trap plants are necessary: with no trap plants, pests destroy the entire crop. But the whole field cannot be devoted to trap plants either, because then no cash crop would remain to harvest. There must therefore be some optimal balance, an intermediate proportion of trap plants that maximizes total farm yield.

In this model, total yield is the yield of an individual cash plant (which depends on the number of pests it hosts) multiplied by the number of cash plants. Holden (2026) showed that in the simplest case where $\lambda = l$, meaning pests disperse from trap and cash plants at the same rate, the optimal number of cash plants in the landscape is,

$$n_C^* = \left[\frac{a - \sqrt{a}}{a - 1}\right] n.$$

The optimal number of trap plants is then simply, $n - n_C^*$. In this chapter we extend that result to allow trap plants to reduce dispersal more effectively than cash plants. To do so let $r = \lambda / l$. By definition, this means, $\lambda = rl$, so $r$ captures the relative dispersal off trap plants relative to dispersal off cash plants. For example, if $r = 0.5$, then pests are half as likely to leave a trap plant as they are to leave a cash plant. If $r = 0.1$, pests are ten times less likely to leave a trap plant than a cash plant.

With this modification the optimal number of cash plants becomes,

$$n_C^* = \left[\frac{a - \sqrt{ra}}{a - r}\right] n.$$

The above expression follows by extending the derivation in Holden (2026): we write total yield as the yield per cash plant multiplied by the number of cash plants, differentiate with respect to $n_C$, set the derivative to zero, and solve for $n_C$. The full algebra mirrors the steps in Holden (2026).

Substituting the above quantity into $n - n_C^*$, gives the optimal number of trap plants,

$$n_T^* = \left[\frac{\sqrt{ra} - r}{a - r}\right] n. \tag{5}$$

To illustrate the results of the optimization, Figure 2 shows total landscape yield as a function of the proportion of trap plants for different levels of the dispersal ratio.

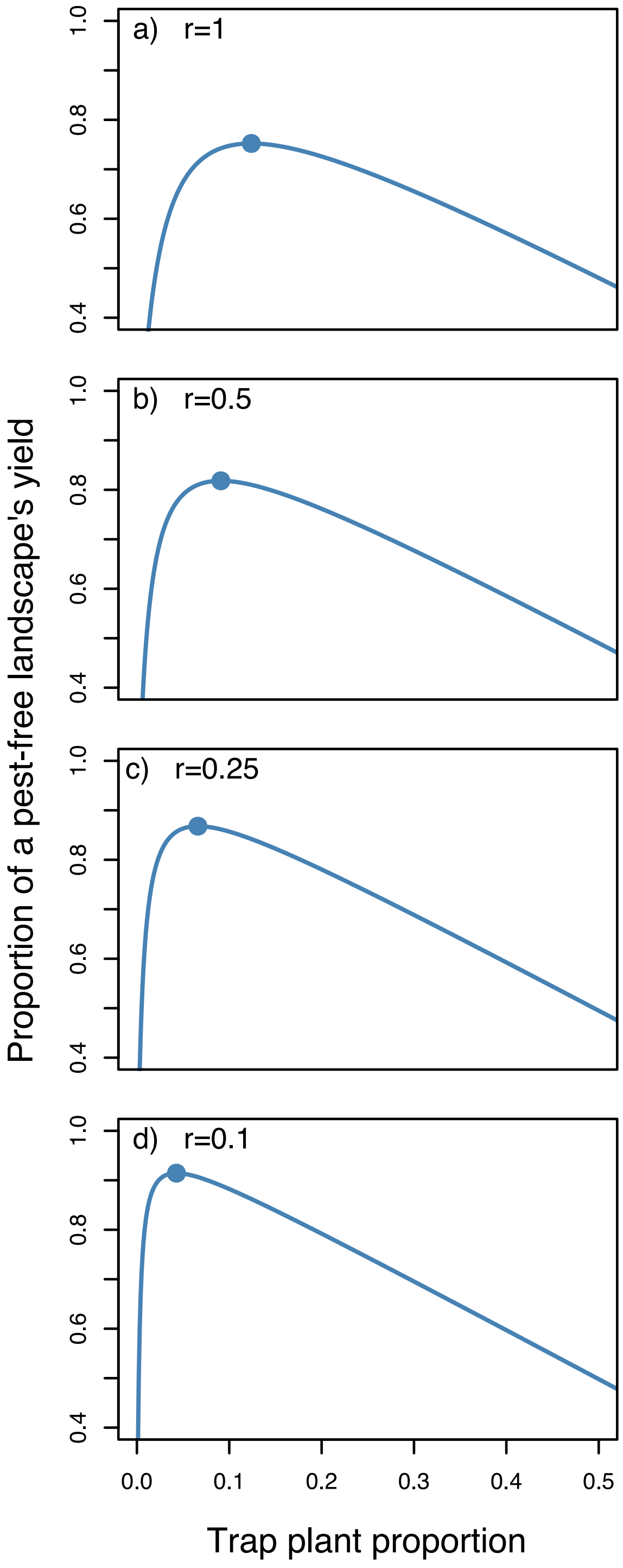

**Figure 2.** Total yield across the whole landscape, scaled relative to the *pest-free cash-crop yield* (the yield that would occur if all plants were cash plants and no pests were present) plotted against the proportion of trap plants. Panels show results for three values of the dispersal ratio, $r$: (a) 1.0 (b) 0.5, (c) 0.25, and (d) 0.1, assuming trap plant attractiveness $a$ = 50. Each line displays the full yield across all possible allocations of trap plants, with the filled circle marking the optimal proportion, as calculated in equation (5). As retention improves (lower $r$), the optimum

shifts toward smaller trap plant proportions and higher maximum yields, highlighting the importance of preventing dispersal from trap plants.

The patterns shown in Figure 2 are especially stark when r = 1, meaning insects are equally likely to disperse from trap plants as from cash plants. The expressions confirm that in this case the optimal number of trap plants makes up an unrealistically high proportion of the landscape. For example, if a trap plant is four times as attractive as a cash plant (a = 4), the grower would have to devote one-third of the landscape to trap plants to maximize yield. Even with a = 10, nearly 25% of the landscape must be trap plants, and with a = 100, still about 10%. These large values, unlikely to be acceptable to growers, provide further rationale for why reducing dispersal from trap plants is critical (see Figure 3).

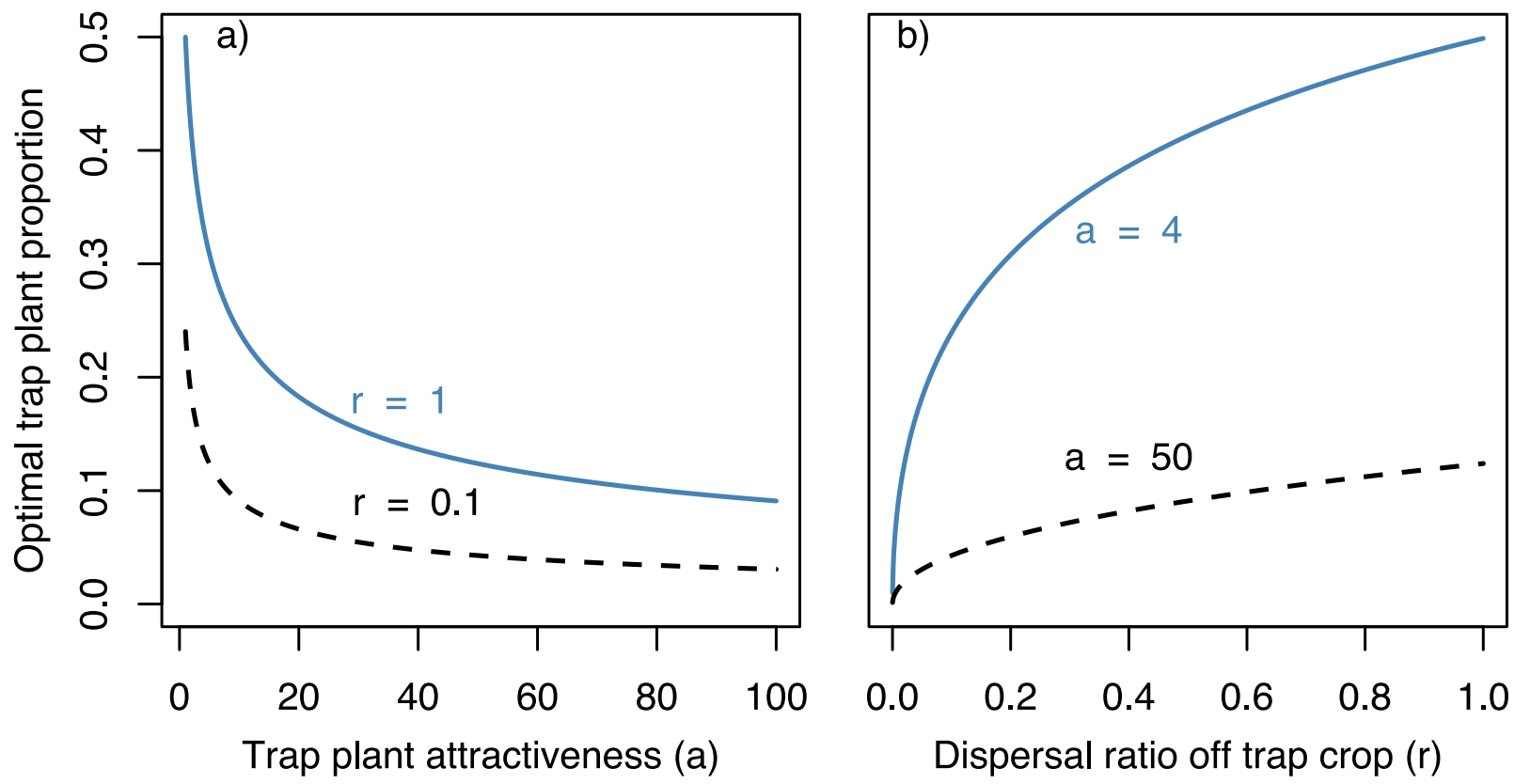

**Figure 3.** Optimal proportion of trap plants in the landscape as a function of (a) trap plant attractiveness (*a*) for two dispersal ratios ($r = 1$ and 0.1), and (b) dispersal ratio (*r*) for two levels of attractiveness ($a = 4$ and 50). Lower values of *r* (greater retention on trap plants) consistently reduce the fraction of land that must be devoted to trap crops, making trap cropping more viable.

The results in Figures 3 and 4 also highlight a critical issue for adoption: even when trap cropping is theoretically effective, the proportion of land required may exceed what growers are willing to sacrifice. In many commercial systems, farmers are reluctant to devote more than about 5–10% of their land to non-cash crops, particularly when margins are tight. Model outputs showing that 20–30% of the landscape must be planted with trap crops therefore point to a major feasibility barrier. This is where agronomic innovation becomes essential. Breeding or selecting trap plants with higher attractiveness (increasing a) or with traits that reduce dispersal from trap plants (lowering r) could reduce the land required. Another practical option is to combine trap cropping with targeted interventions on the trap crop itself (e.g., pesticide application or mechanical removal of pests once aggregated there), effectively boosting retention and making the strategy more commercially viable.

As shown in Figure 3, decreasing the dispersal ratio greatly reduces the number of trap plants required to maximize yield. For example, when r = 1/4, meaning insects are four times less likely to leave a trap plant than a cash plant, the required proportion of trap plants drops to 13.7% when a = 10, and to just 4.8% when a = 100 (compared with 25% and 10% when r = 1). Further reducing r to 0.1, pests ten times less likely to leave a trap plant than a cash plant, lowers the required proportion even more, to about 9% and 3% for a = 10 and a = 100, respectively. This demonstrates how reducing dispersal from trap plants can transform trap cropping from impractical to feasible.

The analysis above considered an extreme case in which pests threaten 100% of yield if no trap crops are planted. In reality, pest damage often reduces yield by only a fraction of the potential harvest. To capture this, we introduce a parameter $\beta$, representing the proportion of yield that is at risk from pests in the absence of trap cropping. This modifies the yield equation, per cash plant, to be,

$$k\left(1-\frac{\beta x^{*} n}{n_{C}}\right).$$

Incorporating $\beta$ into the model allows us to examine how retention influences yield when uncontrolled infestations do not reduce yield by 100 percent (but rather reduce yield by proportion $\beta$) . In this case, the optimal number of trap crops becomes:

$$n_{T}^{*}=\left[\frac{\sqrt{\beta r a}-r}{a-r}\right] n.$$

To illustrate the effect of partial yield loss, we consider two representative cases, before presenting a more complete sensitivity analysis. A low-loss case of $\beta = 0.25$ approximates situations where growers avoid spraying directly on the crop but still engage in other forms of pest management, leading to yield reductions of around 20–30% (see e.g. Oerke 2006; Savary et al. 2019). In contrast, a high-loss case of β = 0.75 represents a severe scenario in which no pest management is undertaken. While extreme, yield losses of this magnitude have been documented in some systems under heavy pest pressure (Oerke 2006; Flood 2010), and they provide a useful upper bound for exploring the potential benefits of trap cropping.

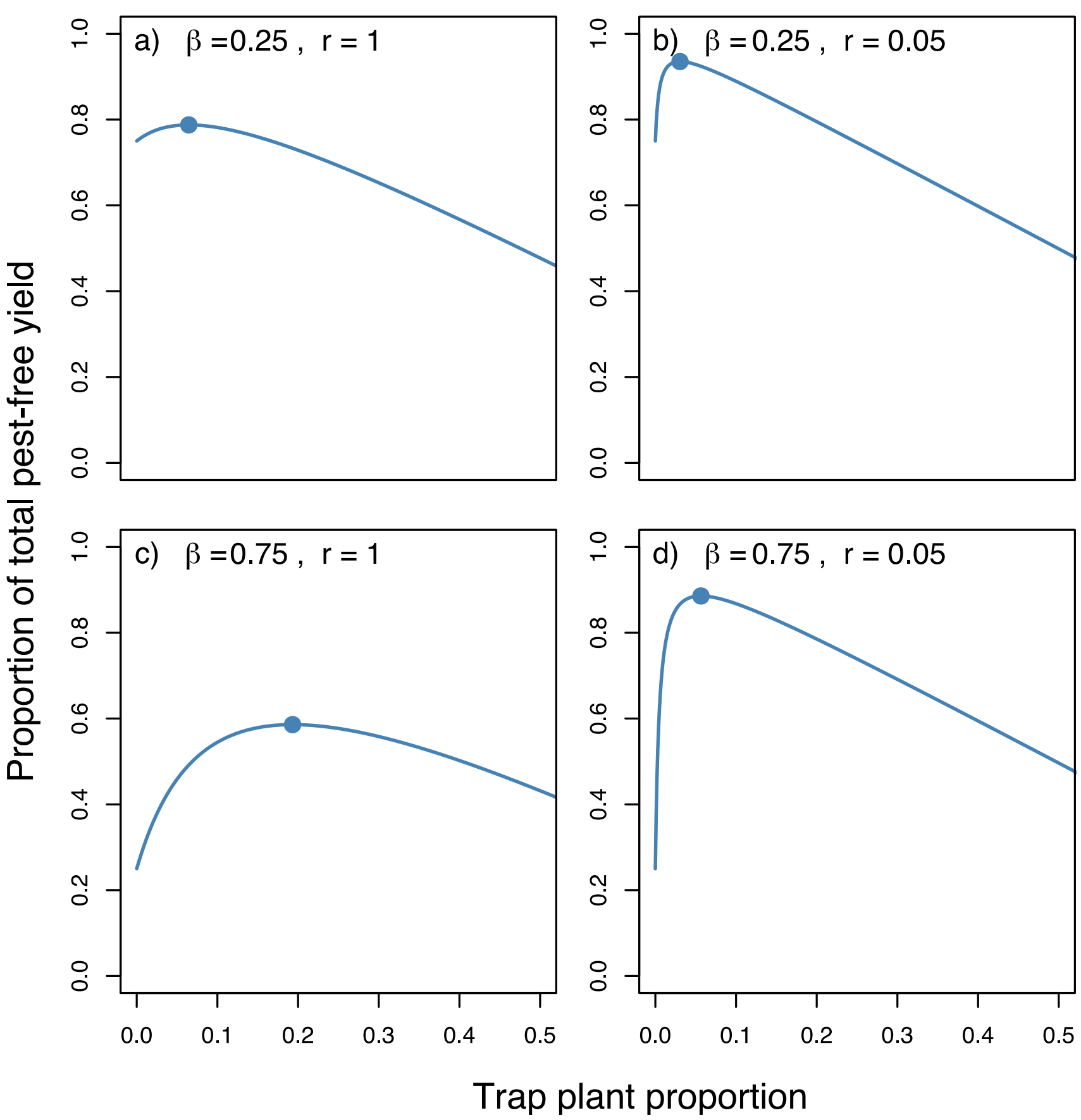

**Figure 4.** Trap cropping is most impactful when yield loss is high and trap plants strongly reduce dispersal. (a-d) Total landscape yield (as a proportion of pest-free yield) plotted against the proportion of trap plants for different levels of yield at risk (β) and retention (r), assuming trap plant attractiveness $a$ = 10. Panels (a) and (b) show results for β = 0.25, meaning only 25% of the yield is vulnerable to pest damage, while panels (c) and (d) show β = 0.75, with 75% of the yield at risk. Within each row, panels (a) and (c), there is no assumed retention advantage on trap plants (r = 1), whereas panels (b) and (d) assume strong retention (r = 0.05). Filled points mark the optimal proportion of trap plants. The results illustrate that when only a small share of yield is threatened (low β), or when trap crops retain pests effectively (low r), the proportion of land required for trap crops to maximize yield is low, making trap cropping more feasible. However, they also show that trap cropping has less impact when the consequences of not trap cropping are low (low yield at risk, panels a and b). Perhaps most strikingly, when yield loss is high, and dispersal off the trap crop is not controlled for (r = 1, c) optimal yield requires very high trap plant proportions.

The results in Figure 4 highlight how partial yield loss and reduced dispersal off trap plants shape the overall yield landscape across different allocations of trap crops. To understand the practical implications for growers, however, it is also important to focus on the optimal proportion of trap plants predicted by the model. Figure 5 shows how this optimal proportion changes with trap plant attractiveness and dispersal ratio. The results emphasize that, even under high-yield-loss scenarios, the optimal proportion of trap plants may remain higher than what most growers would realistically adopt unless trap plants are either extremely attractive or substantially reduce dispersal.

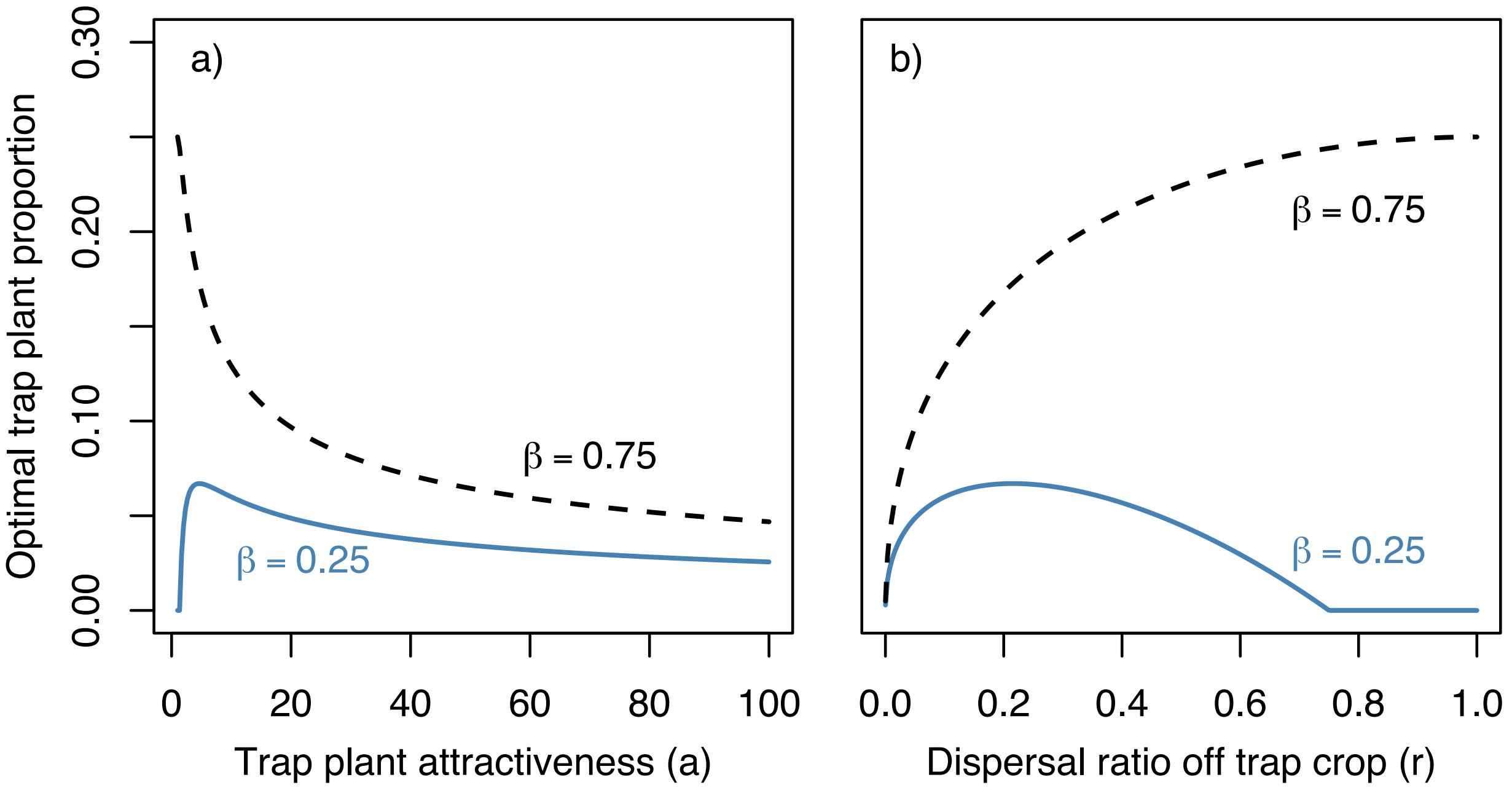

**Figure 5. When a large share of yield is at risk, trap crops must be extremely attractive or substantially reduce dispersal.** (a,b) Optimal proportion of trap plants as a function of trap plant attractiveness (a, panel a) and dispersal ratio off trap plants (r, panel b), shown for two levels of yield at risk ($\beta$). Curves show $\beta = 0.25$ (solid blue, lower-loss case) and $\beta = 0.75$ (dashed black, high-loss case). Baseline parameter values are $r = 1/3$ and $a = 3$, with a varied in panel (a) and r varied in panel (b). For the high-loss case ($\beta = 0.75$), trap plants would need to be at least ~20 times more attractive than cash plants (panel a) or dispersal from trap plants would need to be extremely low (panel b) for the optimal proportion of trap plants to fall below 10% of the field, a level more likely to be acceptable to growers.

The results above focused on identifying the optimal proportion of trap plants. While this is useful for theory, in practice a grower is less concerned with the exact optimum than with the actual yield gains achieved by trap cropping. For example, if allocating 10% of the field to trap plants increases yield only from 0.92 to 0.95 of the pest-free maximum, the benefit may not justify costs. To address this, we return to yield itself as the outcome of interest. Figure 6 generalizes the earlier results by showing how yield depends jointly on attractiveness and dispersal.

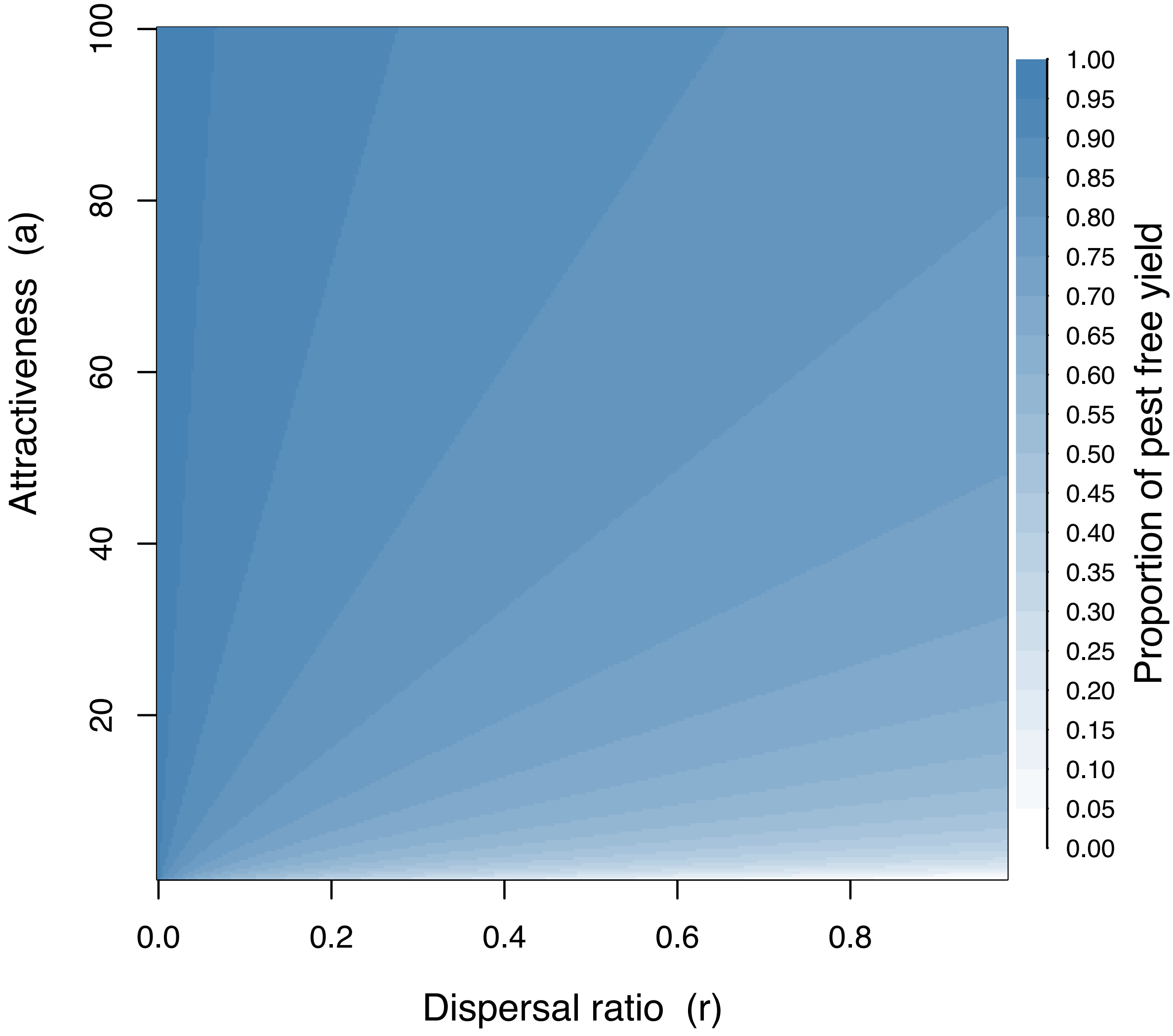

**Figure 6. Yield depends only on the ratio of trap plant attractiveness to dispersal.** Contours show the proportion of pest-free yield achieved across combinations of trap plant attractiveness ($a$) and dispersal ratio off trap plants ($r$), assuming all yield is at risk in the absence of control ($\beta = 1$). The straight contour lines through the origin indicate that yield is determined entirely by the ratio $r/a$. Thus, reducing dispersal from trap plants or increasing their attractiveness improves yield equally when scaled by the same factor. This result generalizes the earlier analyses: preventing dispersal is just as important as attracting pests in determining the practical benefits of trap cropping.

The patterns in Figure 5 already suggested that trap cropping only becomes practical when dispersal is very low or attractiveness very high. Figure 6 explains why: the contours of equal yield are straight lines through the origin, showing that yield depends only on the ratio of dispersal to attractiveness (r/a). This means increasing attractiveness or decreasing dispersal has an equivalent effect when scaled by the same factor, generalizing the inverse importance hinted at in the previous figure.

Mathematically, this can be shown by expressing yield as

$$p\left(1-\frac{\beta z}{1-p+pz}\right),$$

where $z = r/a$ and $p = n_C/n$. For any fixed maximum yield loss (β) and crop allocation (p), yield is therefore a function of $r/a$. Put simply, a grower can achieve the same yield either by increasing attractiveness or by reducing dispersal off trap plants by an equivalent factor. This generalizes the earlier results: preventing dispersal is just as important as attracting pests in the first place. Crucially, it also allows growers to evaluate whether the gains from trap cropping are meaningful in practice, since the key question is not whether a fixed number of trap plants is 'optimal,' but whether the resulting yield improvement is large enough to warrant allocating the required land to trap crops.

**Discussion**

Our results reinforce a key conclusion from Holden et al. (2012): Preventing pests from dispersing off trap plants back into the cash crop is critical for trap cropping to succeed at commercial scales. We add nuance to that conclusion by showing that the success of trap cropping depends jointly on two interacting features — how attractive trap plants are, and how much they reduce pest dispersal compared to the cash crop. In our formulation, long-term pest densities depend on the combined effect of these two factors together with how much of the field is devoted to trap plants. When either attraction is weak or dispersal from the trap crop is high, trap cropping only works if a very large proportion of the field, often more than 20 percent, is planted with trap plants.

The analysis shows that both attraction and reduced dispersal are equally powerful tools for shifting pests off the cash crop. Increasing both how strongly pests are drawn to trap plants and decreasing how readily they leave them produce the same overall reduction in pest density. This finding clarifies why previous modelling studies, such as those by Banks and Ekbom (1999) and Potting and Perry (2005), achieved success only when trap crops occupied large fractions of the field: without reduced dispersal, even very attractive trap plants could not keep pests away from the main crop for long. Conversely, small reductions in dispersal from trap plants can sharply reduce the land required for trap crops—from around one-quarter of the field to as little as five per cent in some scenarios—making the strategy much more feasible.

Holden et al. (2012) interpreted their results to suggest that retention (the opposite of dispersal) was more important than attraction. Our results suggest instead that the two factors are mechanistically equivalent: both operate through the same underlying trade-off between how many pests reach the trap crop and how many stay there. The earlier emphasis on retention arose mainly from how the parameters were defined, not from an intrinsic difference in importance. When both are placed on a common scale, they act as equal levers for reducing pest pressure on the cash crop.

In practice, attraction is often constrained by the biological traits of the plants chosen. Once a trap–cash plant pair is identified, it is hard to alter how strongly the pest prefers one over the other. By contrast, dispersal from trap plants can be influenced through management. Growers can reduce backflow of pests into the cash crop by treating trap plants with selective insecticides, physically removing insects through vacuuming or sticky panels, or by placing traps or barriers in configurations that hinder pest movement (e.g. Moreau & Isman 2011, Swezey, Nieto & Bryer 2014). These interventions increase the amount of time pests remain on the trap crop, enhancing its effectiveness.

The trade-off for growers is straightforward: every trap plant occupies land that could otherwise produce revenue via cash plants that produce harvestable yield. Our yield-optimisation analysis captures this balance explicitly. As the proportion of trap plants increases, yield rises initially because pest damage falls, but beyond a certain point the loss of productive area outweighs the pest-control benefit. The optimum occurs where these two effects balance.

The model shows that when pests disperse freely from trap plants, the optimum proportion of trap plants is often unrealistically high, typically 20–30 per cent of the field for moderately attractive trap plants, making adoption uneconomical. However, even modest reductions in dispersal can shift this optimum dramatically. When pests are several times less likely to leave a trap plant than a cash plant, the required trap area can fall to around 5 per cent, a level that is both agronomically and economically feasible. This pattern explains why trap cropping has rarely succeeded in large-scale commercial systems without additional interventions to prevent backflow.

By linking pest dynamics to yield, the optimisation framework helps clarify what "effective" means in practical terms: not simply reducing pest density but doing so enough to raise total yield after accounting for land sacrificed to trap plants. Our analysis shows that without strong retention, even quite attractive trap crops would demand 20–30% of the field be devoted to trap plants to optimise yield, well beyond the 5–10% many growers are willing to sacrifice. Only when trap crops are either extremely attractive or substantially reduce dispersal does this ratio approach levels a grower would realistically consider.

These modelling insights align closely with empirical evidence. Holden et al. (2012) reviewed dozens of trap-cropping systems and showed that every commercial-scale success paired attraction with some mechanism of pest arrestment: pesticide applications to trap plants (e.g. Srinivasan & Krishna Moorthy 1991; Dogramaci et al. 2004; Lu et al. 2009), physical removal such as vacuuming (Swezey, Nieto & Bryer 2014), sticky traps (Moreau & Isman 2011), or planting trap crops at exceptionally high densities (Michaud, Qureshi & Grant 2007). In contrast, systems that relied on attraction alone generally failed. More recent studies have begun explicitly quantifying retention alongside attraction. For example, Blaauw et al. (2017) measured how long *Halyomorpha halys* adults remained on sunflower trap crops relative to cash crops; Boiteau et al. (2014) evaluated arrestment of *Colorado potato beetle* on solanaceous trap plants; and Wallingford et al. (2018) tested oviposition retention of *Drosophila suzukii*. These studies illustrate both the variability of retention across pest-plant systems and its central role in determining trap-cropping efficacy. Our framework helps interpret such results mechanistically, providing a quantitative rationale for why reducing dispersal from trap plants correlates so strongly with field success.

Retention need not rely solely on plant traits. Our results highlight how complementary tactics can effectively reduce dispersal from trap plants back into the main crop. Banks & Laubmeier (2024) demonstrated through coupled pest–enemy models that when predators or parasitoids tracked pest aggregations, pest densities fell by more than half. In our framework, this effect is equivalent to reducing dispersal off the trap crop. Similarly, applying selective insecticides to trap crops, or using physical controls such as vacuuming or sticky traps, all function to prevent dispersal off the trap crop as well. This suggests that trap cropping should be viewed not as a stand-alone strategy but as one component of integrated pest management (Vinatier et al. 2012, Angon et al. 2023). The greatest potential lies in combining attractive trap plants with additional interventions that ensure pests remain on trap plants long enough to be killed or removed.

Our analysis simplifies many aspects of real systems. It assumes pests move freely among plants, that all plants are identical within their type, and that yield loss is proportional to pest density. In practice, yield–pest relationships are often nonlinear and vary widely among crops, pest species, and crop growth stages (Pedigo et al. 1986; Oerke 2006). Such nonlinear relationships may alter the optimal proportion of trap plants. Field conditions add further complexity: spatial layout, timing of crop development, and pest life history all influence outcomes. Moreover, growers care more about cumulative pest pressure during a season than about long-term equilibria. Our analytical results approximate cumulative pest load well when dispersal equilibrates quickly, but further work is needed to confirm when this holds empirically in field conditions.

Future directions include testing our predictions in spatially explicit or stochastic models, incorporating harvest timing and economic thresholds, and parameterising models with empirical data on dispersal and attraction. Breeding or selecting trap varieties that combine high attraction with low dispersal remains a promising avenue, as does integrating trap crops with biological control agents that naturally suppress pest escape.

Overall, our findings sharpen a principle that has been implicit in the trap-cropping literature for decades: the success or failure of trap cropping hinges on preventing pest dispersal off the trap crop. Attraction determines whether pests arrive; reduced dispersal determines whether they stay. By reformulating existing models and extending them to yield optimisation, we show why controlling dispersal off trap plants so decisively shifts trap cropping from impractical to feasible. This insight moves the design challenge away from simply finding "the most attractive" plant and toward strategies that ensure pests remain on trap plants once they arrive, whether by plant traits, natural enemies, or complementary management tactics. Focusing on this principle will improve the likelihood that trap cropping contributes meaningfully to sustainable pest management at commercial scales.

**Acknowledgements**
I am grateful to Wopke van der Werf, whose insightful review of Holden et al. (2012) first raised the idea of the odds ratio explored in this chapter. I only came to fully appreciate the depth of his insight while developing the optimization problem presented in this work.